\documentclass[prb,showpacs,nofootinbib,twocolumn,
superscriptaddress,preprintnumbers,amsmath,amssymb]{revtex4}
\usepackage{graphicx}
\usepackage{amsbsy}
\usepackage{dcolumn}
\usepackage{bm}
\begin{document}
\title{Vortex lattice disorder in  \YBCO{7-\delta} probed using 
  \bnmr\ }
\author{H. Saadaoui}
\affiliation{Department of Physics and Astronomy, 
  University of British Columbia, Vancouver, BC, Canada, V6T 1Z1}
\author{W.~A. MacFarlane}
\affiliation{Chemistry Department, 
  University of British Columbia, Vancouver, BC, Canada, V6T 1Z1}
\author{Z. Salman}
\altaffiliation[Present address: ]{LMU PSI, Villigen CH.}
\affiliation{TRIUMF, 4004 Wesbrook Mall, Vancouver, BC, Canada, V6T 2A3}
\author{G.~D. Morris}
\affiliation{TRIUMF, 4004 Wesbrook Mall, Vancouver, BC, Canada, V6T 2A3}
\author{Q. Song}
\affiliation{Department of Physics and Astronomy, 
  University of British Columbia, Vancouver, BC, Canada, V6T 1Z1}
\author{K.~H. Chow}
\affiliation{Department of Physics, 
  University of Alberta, Edmonton, AB, Canada, T6G 2G7}
\author{M.~D. Hossain}
\affiliation{Department of Physics and Astronomy, 
  University of British Columbia, Vancouver, BC, Canada, V6T 1Z1}
\author{C. D. P. Levy}
\affiliation{TRIUMF, 4004 Wesbrook Mall, Vancouver, BC, Canada, V6T 2A3}
\author{A.~I. Mansour}
\affiliation{Department of Physics,
  University of Alberta, Edmonton, AB, Canada, T6G 2G7}
\author{T.~J. Parolin}
\affiliation{Chemistry Department, 
  University of British Columbia, Vancouver, BC, Canada, V6T 1Z1}
\author{M. R. Pearson}
\affiliation{TRIUMF, 4004 Wesbrook Mall, Vancouver, BC, Canada, V6T 2A3}
\author{M. Smadella}
\affiliation{Department of Physics and Astronomy, 
  University of British Columbia, Vancouver, BC, Canada, V6T 1Z1}
\author{Q. Song}
\affiliation{Department of Physics and Astronomy, 
  University of British Columbia, Vancouver, BC, Canada, V6T 1Z1}
\author{D. Wang}
\affiliation{Department of Physics and Astronomy, 
  University of British Columbia, Vancouver, BC, Canada, V6T 1Z1}
\author{R.~F. Kiefl}
\affiliation{Department of Physics and Astronomy, 
  University of British Columbia, Vancouver, BC, Canada, V6T 1Z1}
\affiliation{TRIUMF, 4004 Wesbrook Mall, Vancouver, BC, Canada, V6T 2A3}
\affiliation{Canadian Institute
for Advanced Research, Toronto, Canada M5G 1Z8}
\newcommand{\eli}{$^8$Li}
\newcommand{\elip}{$^8$Li$^+$}
\newcommand{\ebe}{$^{11}$Be}
\newcommand{\cs}{C$_{60}$}
\newcommand{\Lpd}{$\lambda_{\rm L}$}
\newcommand{\BCO}[1]{Ba$_{2}$Cu$_{3}$O$_{{#1}}$}
\newcommand{\YBCO}[1]{Y\BCO{#1}}
\newcommand{\YBCOd}{YBa$_{2}$Cu$_{3}$O$_{7-\delta}$}
\newcommand{\BISCO}{Bi$_2$Sr$_2$Cu$_2$O$_{8+\delta}$}
\newcommand{\STO}{SrTiO$_{3}$}
\newcommand{\PCCO}{Pr$_{2-x}$Ce$_x$CuO$_4$}
\newcommand{\PCCOd}{Pr$_{2-x}$Ce$_x$CuO$_{4-\delta}$}
\newcommand{\CO}{$\rm CuO_2$}
\newcommand{\Li}{${}^8$Li$^+$}
\newcommand{\NbSe}{NbSe$_2$}

\newcommand{\msr}{$\mu$SR}
\newcommand{\lem}{LE-$\mu$SR}
\newcommand{\bnmr}{$\beta$-NMR}

 \newcommand{\equ}[2]{\begin{equation}\label{#1}{#2}\end{equation}}
\newcommand{\meq}[2]{\begin{eqnarray}\label{#1}{#2}\end{eqnarray}}
\newcommand{\Tc}{$T_c$}
\newcommand{\bq}{{\bf q}}
\newcommand{\bk}{{\bf k}}
\newcommand{\bG}{{\bf G}}
\newcommand{\bg}{{\bf g}}
\newcommand{\bkp}{{\bf k'}}
\newcommand{\br}{{\bf r}}
\newcommand{\bR}{{\bf R}}
\newcommand{\bp}{{\bf p}}
\newcommand{\etal}{{\it et al.}}
\begin{abstract}
  $\beta$-detected NMR (\bnmr) has been used to study  vortex lattice
  disorder near  the surface   of the high-$T_C$ superconductor
  \YBCOd\ (YBCO).  The magnetic field distribution from the  vortex
  lattice  was detected by implanting a low energy beam of highly polarized
  $^8$Li$^+$ into a  thin overlayer  of silver on 
  optimally doped, twinned and detwinned YBCO samples.  The resonance
  in Ag broadens significantly below the transition temperature $T_C$ 
  as expected from the emerging field lines of   
  the vortex lattice in YBCO.
  However, the lineshape is more symmetric and the dependence on the applied
  magnetic field is much weaker than expected from an ideal vortex lattice, 
  indicating that the vortex density varies across the face of the
  sample, likely due to  pinning at twin boundaries.  
  At low temperatures the broadening from such disorder
  does not scale with the superfluid density.
\end{abstract}
\maketitle

\section{Introduction}
The vortex state of  cuprate superconductors is of central importance in
understanding  high-$T_C$ superconductivity (HTSC). One the most well studied
quantities  is the internal  magnetic field distribution $p(B)$  associated
with the vortex lattice 
(VL).\cite{BrandtJLTP05,SalmanPRL07,KhasanovPRL07,SonierPRB07} 
As discussed below, several  methods can be used to measure  $p(B)$, which
depends on the  London penetration depth $\lambda$, 
the coherence length $\xi$,\cite{BrandtJLTP88,Belousov03} 
and, to a lesser extent, the internal
structure of the vortices,\cite{MillerPRL02} and non-linear and non-local
effects.\cite{SonierPRB07,AminPRB98,SonierPRL99} The form of $p(B)$  has a
distinctive asymmetric  shape due to the spatial magnetic inhomogeneity
  characteristic of an ordered  two-dimensional (2D) lattice of vortices. 
One basic feature in $p(B)$ is a prominent high field tail
associated with the vortex cores, which depends on the magnitude of $\xi$.  
There is also a saddle point in the local field profile located   between two
vortices. This gives rise to a Van Hove singularity or sharp peak in
$p(B)$ below  the average  field. The overall width or second
moment of $p(B)$ depends primarily  on  the London penetration depth
$\lambda$, the lengthscale over which the magnetic field is screened.
Anisotropy of the Fermi surface or the superconducting order
parameter can result in a different  VL but the main
features are similar for any ordered lattice.\cite{FranzPRB96,AminPRL00}

Another general feature associated with any real VL is disorder
arising from vortex pinning at  structural defects and
impurities.\cite{HilgenkampRMP02,BlatterRMP94} Structural defects  are
present in all superconductors to some degree, but may be more 
prevalent  in structurally complex compounds such as YBCO. For example,
YBCO's slightly orthorhombic structure facilitates crystal twinning, i.e. 
in a single crystal, there are generally domains with the
nearly equal $a$ and $b$ directions interchanged.
Separating such twin domains are well defined 45$^{\circ}$ grain boundaries
or twin boundaries which have been shown to be effective extended vortex
pinning sites.\cite{BishopSc92,YethirajPRL93,GammelPRL87,HerbsommerPRB00}
Scanning tunneling microscopy (STM) imaging of a twinned YBCO
crystal show that the areal vortex density is strongly modified  
by the twin boundaries.\cite{Maggio-Aprile97}  Small-angle neutron scattering 
(SANS) studies of YBCO confirm that the twin boundaries strongly deform the
VL.\cite{YethirajPRL93,SimonPRB04}   Understanding the influence
of such  structural defects on the VL 
has been the subject of intense theoretical 
work,\cite{FisherPRB91,NelsonPRL92,MikitikPRB09} and 
is important for two main reasons. 
Firstly, it affects $p(B)$ and thus adds  uncertainty to measurements
of fundamental quantities like $\lambda$ and $\xi$,  since it can be
difficult to isolate such extrinsic effects from changes in fundamental
quantities of interest. Secondly,  the degree of pinning of vortices
determines the critical current density which is important for many
applications.\cite{Bartolome08} 

Measurements of the vortex state field distribution $p(B)$ are most
often done using SANS,\cite{YethirajPRL93,SimonPRB04}
nuclear magnetic resonance (NMR),\cite{nmrpapers} and conventional  muon-spin
rotation (\msr).\cite{SonierRMP00}  All these methods  probe the
VL in the bulk and can be applied over a wide range of magnetic
fields.  It is also possible to probe the  magnetic field
distribution $p(B)$ near the surface of the sample  using low energy-\msr\
(LE-\msr) in low magnetic fields.\cite{Morenzoni99} 
Recently we have demonstrated that similar information on
$p(B)$ near a surface can be obtained using \bnmr.\cite{SalmanPRL07}
This has the advantage that it can be applied over a wide
range of magnetic fields.

In this paper, we report  measurements of the VL above  the
surface of the cuprate superconductor \YBCOd\  
using \bnmr.\cite{KieflPC03,MorrisPRL04,SalmanPRB04,SalmanNl07,
  XuJMR08,SaadaouiPB09-V} 
The \Li\ beam was implanted into a  thin silver overlayer evaporated onto
several  YBCO samples. Measuring in the Ag allows one to isolate  the
contribution to $p(B)$ from {\it long wavelength disorder}, i.e. 
disorder that occurs on length scales much 
longer than the vortex spacing and $\lambda$ due to
structural defects such as twin and grain boundaries. This is possible because
the field distribution broadening just outside the superconductor due to the
VL inside has a
very distinctive field dependence. In particular, it vanishes in high magnetic
fields where the VL  spacing  becomes less than the characteristic distance of
the probe from the superconductor. On the other hand, long wavelength
disorder  has a much weaker dependence on magnetic field and
dominates the observed $p(B)$ in the high field limit.   Our results show
evidence for significant broadening of $p(B)$  from such long wavelength
disorder on  the scale of  $ D \approx\ 1~\mu m$, which is attributed to
pinning at twin or other grain boundaries. The magnitude of the broadening is
similar to that  observed in  bulk \msr\ measurements, suggesting that the
same broadening contributes to $p(B)$ in  bulk \msr\ measurements. 
There is a crossover such that near $T_C$, where $\lambda\gg D$,
the broadening scales with the superfluid density, whereas  at lower
temperatures, where $\lambda\ll D$,  the broadening does not track the
superfluid density. We discuss the consequences of this for the
inference of $\lambda(T)$ from measurements of $p(B)$ 
in polycrystalline superconductors.

The paper is organized as follows: 
section II reviews the theory for the  field distribution, and its second
moment near the surface  of a  superconductor. Section III contains all  the
experimental details. In  Section IV, we present the results. 
Finally in section V we discuss the results and draw conclusions.

\section{The magnetic field distribution $p(B)$ in the vortex state}
In a type II superconductor, above the lower critical field $B_{c1}$, 
the magnetic field penetrates the sample {\it inhomogeneously} forming a
lattice of magnetic vortices,  each carrying  a flux quantum, $\Phi_0
= h/2e$. In a perfect crystal, intervortex interactions lead to a long-range
ordered 2D lattice of vortices, usually of triangular (hexadic)
symmetry.\cite{BrandtJLTP05} At the core of each vortex (a cylinder of radius
approximately the superconductor's  coherence length $\xi$), the local
magnetic field is maximal. Outside the core, concentric circulating
supercurrents partially screen the field which thus falls exponentially 
with a lengthscale $\lambda$. The average magnetic field in the VL is the
applied field $B_{0}$ for flat samples, where demagnetization effects 
are negligible.\cite{PooleBK95,pozekPC96,SteegmansPC98}  At a given field, the
average vortex spacing, i.e. the lattice constant of the VL, $a$ is fixed. 
For the triangular lattice this is
\begin{eqnarray}\label{a-vl}
  a=\sqrt{\frac{2\Phi_0}{\sqrt{3}B_0}}
  \approx\frac{1546\mbox{~nm}}{\sqrt{B_0({\rm mT})}}.
\end{eqnarray}
If one considers the profile of the magnetic field along a 
line in the lattice (perpendicular to the direction of the applied field), it
is thus {\it corrugated} with a period determined by $a$.
This inhomogeneity in the magnetic field causes a 
characteristic broadening in local magnetic resonance probes
such as the muons in $\mu$SR or the host nuclei in NMR.
Since  the muon (or host nuclear spin) is  at  a well-defined 
lattice site(s), it  samples the VL  with a  grid spacing given by the lattice
constant of the crystal. Since this is much smaller than the VL constant, the
resulting field distribution $p(B)$ provides a random sampling  of the
spatially inhomogeneous field $B(r)$  over the  VL unit cell:
\begin{eqnarray}\label{pB-def}
p(B) = \frac{1}{A} \int_A \delta(B-B(r)) dr,
\end{eqnarray}
where the integral is over a unit cell of the VL of area $A$.
In this paper we are concerned with  the $z$-component  of the magnetic
field $B_z$ (parallel to the $c$-axis of YBCO samples), 
and refer to it  simply as $B$.

For an ideal triangular VL,  the spatial
dependence of  the $z$-component of the magnetic field  
in or outside a type II superconductor
follows the modified London equation,
\begin{equation}\label{br-london-b1}
-\nabla^2 B -\frac{\partial^2B}{\partial z^2} 
+\frac{ B}{\lambda^2} \Theta(z)=\frac{\Phi_0}{\lambda^2} \Theta(z) \sum_{\bR}
\delta({\br}- \bR). 
\end{equation}
Here $\lambda=\lambda_{ab}$ when the applied field is along the $c$-axis 
and  the screening supercurrents flow in the ab plane, 
$\nabla^2$ is the 2D Laplacian, $\Theta(z)=1$
for $z>0$ and  zero otherwise, 
${\bf r}$ is a 2D vector in the xy plane, 
${\bf R}$ are the Bravais lattice vectors for the VL.
We define the $z$ axis as the normal to the surface 
of a superconducting slab with negative $z$ outside the superconductor.
The solution of Eq. (\ref{br-london-b1}) is easily obtained using the Fourier
transform, $B(\br,z) = B_0 \sum_{\bk}e^{i\bk\cdot \br}F(\bk,z)$,  where 
the dimensionless Fourier components, $F(\bk,z)$, 
are given by,\cite{NiedermayerPRL99} 
\equ{FourierB}{
  F(\bk,z)
  =\frac{1}{\lambda^2}\Big[\frac{\Theta(-z)e^{kz}}{\Lambda(\Lambda +k)}
  +\frac{\Theta(z)}{\Lambda^2}(1-\frac{k}{\Lambda +k}e^{-\Lambda z})\Big].}
Here $\Lambda^2=k^2+\frac{1}{\lambda^2}$,
and $\bk= \frac{2\pi}{a}[n{\bf x}+\frac{2m-n}{\sqrt{3}}{\bf y}]$ 
are the reciprocal lattice vectors of the triangular VL, where $m,n=0,\pm
  1,\pm 2...$. A cutoff function $C(k)$, approximated by a simple  
Gaussian $C(k)\approx e^{-\frac{\xi^2k^2}{2}}$,  
can be used to account for the finite size of the vortex core, 
where $F(\bk,z)$ is replaced by $F(\bk,z)C(k)$.\cite{SalmanPRL07,SonierPRB07}
However, the corrections due to $C(k)$ are very small in our case, so it
will be omitted. An approximate solution for the magnetic field along $z$
(both inside and outside the superconductor) is given by  
\equ{Bzr_eq}
{B({\bf r},z)=B_0\sum_{\bk}F(\bk,z)\cos(\bk\cdot \br).}
The second moment of $p(B)$ at a depth $z$,  $\sigma^2 =
\langle B^2 \rangle -\langle B \rangle^2$, where $\langle..\rangle$ is the
spatial average, is given by
\equ{scm-vl}
{\sigma^2=B^2_0\sum_{\bk\neq 0}F^2(\bk,z).}

\begin{figure}
  \includegraphics[width=\columnwidth]{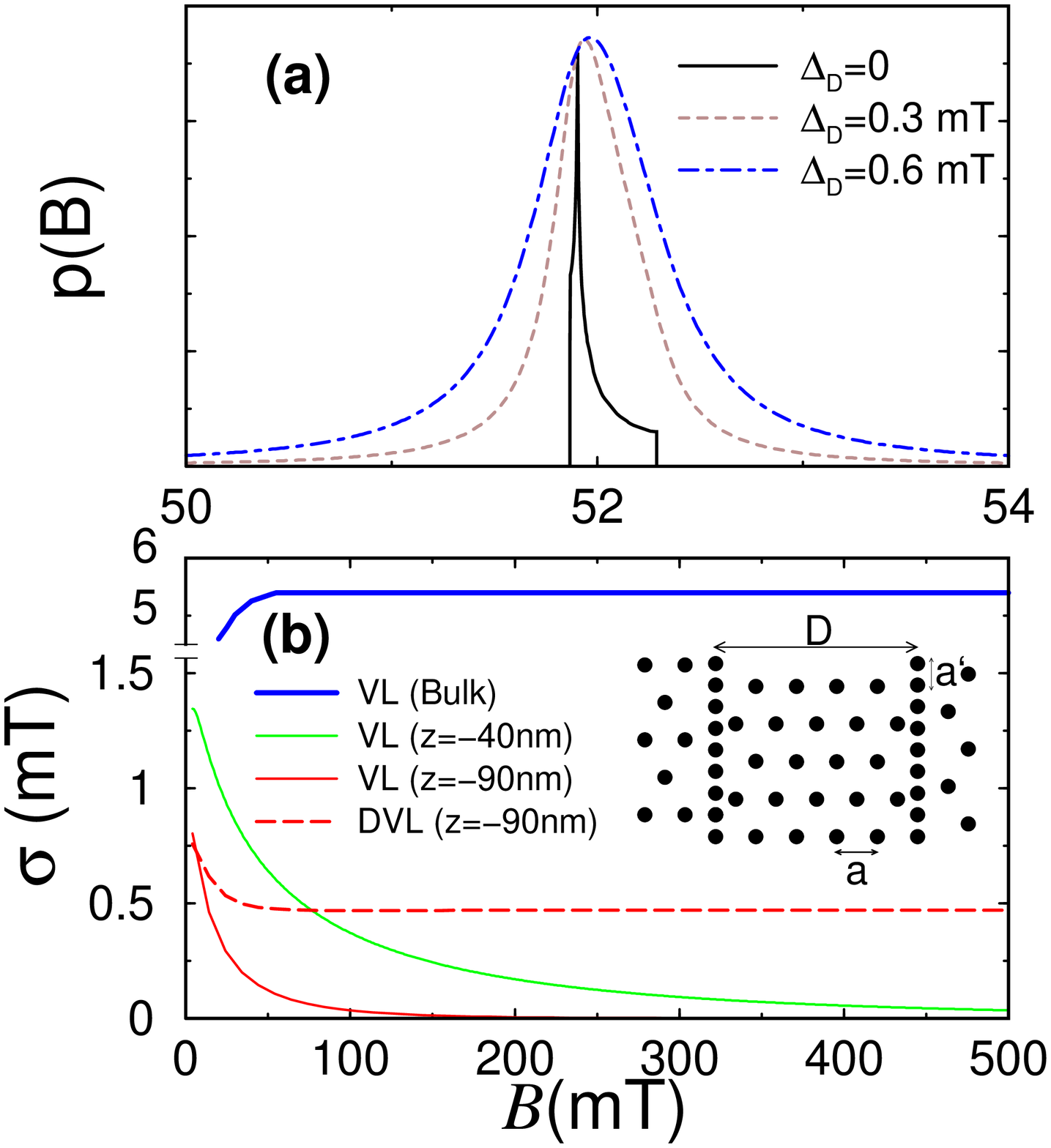}
  \includegraphics[width=\columnwidth]{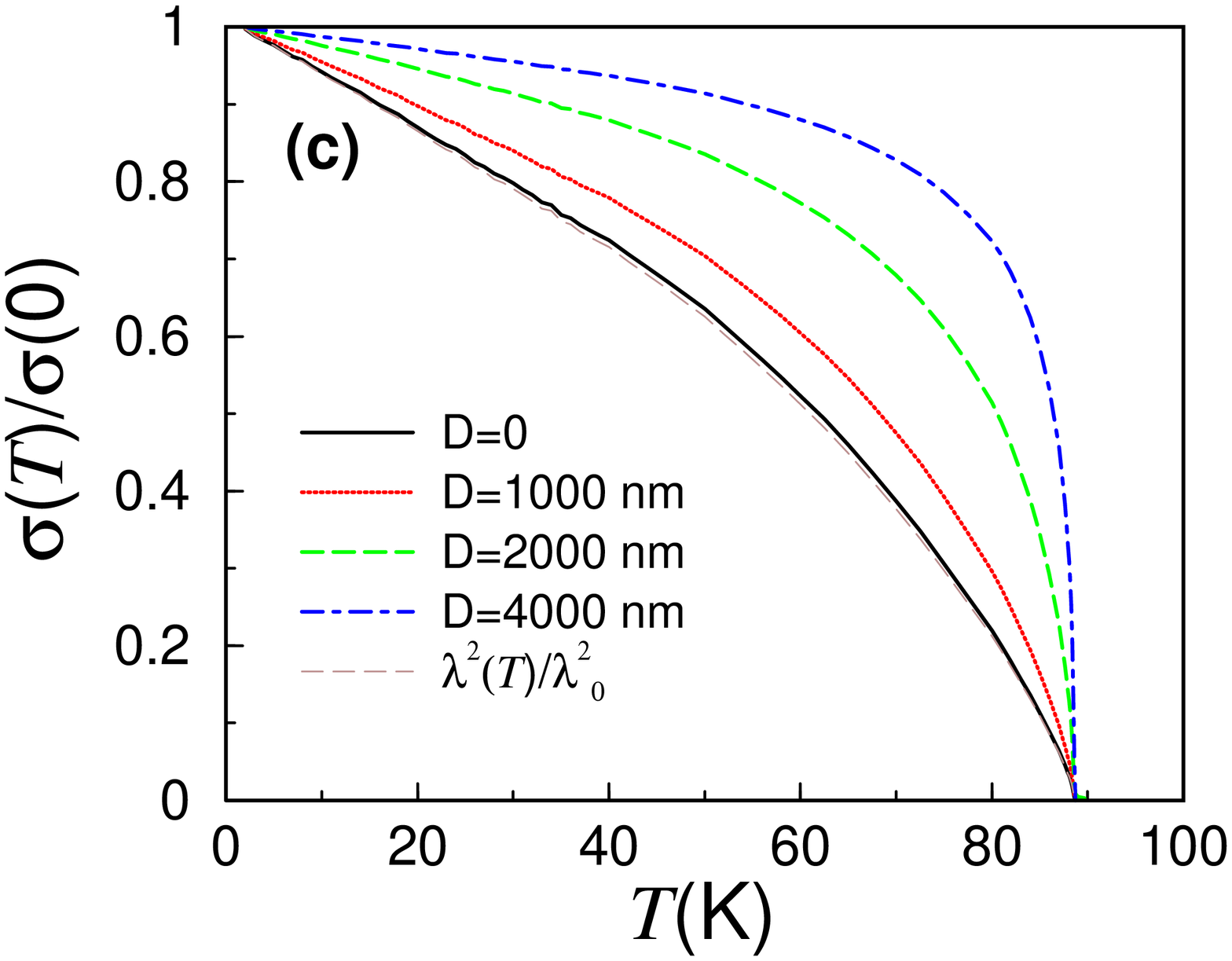}
  \caption{(Color online) (a) Simulation of $p(B)$ in an applied field of 52 mT
    at a distance 90 nm  from the superconductor using Eq. (\ref{Bzr_eq})   
    convoluted with a Lorentzian of width $\Delta_{\rm D}$. 
    (b) The broadening of $p(B)$ versus the applied field. Solid lines
    represent $\sigma$ from Eq. (\ref{scm-vl}) in the bulk, and  
    40 and 90 nm away from the superconductor. Long dashed line shows 
    $\sigma$ at 90 nm from Eq. (\ref{scm-dvl}) for $D=4$ $\mu$m and $f=0.1$.
    Inset: sketch of a possible vortex arrangement including vortex trapping
    at twin boundaries spaced by $D$ and a regular triangular vortex
    lattice elsewhere. (c) $\sigma(T)$ (normalized at $T=0$), from
    Eq. (\ref{scm-dvl}) for     $B_0=52$ mT, $z=-90$ nm, and $f=0.1$, is
    plotted against $T/T_C$ for different      values of $D$.     A $d$-wave
    temperature dependence of      $\lambda(T)$ is used.\cite{Bonn} 
    In all figures 
    $\lambda(0)=150$ nm.} 
  \label{simfig}
\end{figure}
The field distribution for a perfectly ordered  
triangular VL calculated from Eqs. (\ref{pB-def}) and (\ref{Bzr_eq}),
at $B_0=52$ mT and outside the superconductor at $z=-90$ nm  
and $\lambda=150$ nm (relevant to YBCO at $T \ll T_C$),
is presented in Fig. \ref{simfig}(a) ($\Delta_{\rm D}=0$, defined below). 
It shows the characteristic 
high field skewness with a cutoff corresponding to the field
at the core of the vortices. The sharp peak corresponds to the 
most probable field $B_{sad}$ at saddle points in $B(r)$
midway between adjacent vortices.  The low field cutoff 
occurs at the center of an elementary triangle of vortices.
As we move farther from the superconductor, $B_{sad}$ moves towards the 
applied field as the field approaches uniformity for $z\to -\infty$, 
$i.e.$, $p(B)\to \delta(B-B_{0})$. This crossover occurs as
$\exp(\frac{2\pi}{a}z)$, as the $z$ variation of the Fourier components
$F(\bk,z)$ in Eq. (\ref{FourierB}) is controlled by $k$ which 
takes values equal to or larger than $2\pi/a$.
However, if instead we consider a simple non-superconducting 
overlayer instead of free space, then the limiting $p(B)$ will be
the intrinsic lineshape in the overlayer material. 

As mentioned above $p(B)$ is also affected by  disorder in the VL due to
pinning at structural defects in the crystal, where the
superconducting order parameter is suppressed. Such disorder causes broadening
of the magnetic resonance, obscuring  the features expected from an ideal
VL,\cite{SonierRMP00} adding  uncertainty to parameters of interest such
as  $\lambda$ and $\xi$. Relatively little is known about the detailed
characteristics of this disorder.  Accounting for the disorder of the VL is
most often  done by smearing the ideal lineshape with a Gaussian or Lorentzian
distribution of width $\Delta_{\rm D}$, where the latter is a phenomenological
measure of the degree  of disorder.\cite{BrandtJLTP88,HarshmanPRB93}
Calculated distributions  for an applied field ${B_0} = 52$~mT  are shown in
Fig. \ref{simfig}(a) for $\Delta_{\rm D}=$ 0.3 and 0.6 mT, 
together with the ideally ordered VL ($\Delta_{\rm D}=0$). 
Such disorder is more pronounced outside the
superconductor and renders the lineshape symmetric
when the depth dependent intrinsic VL broadening
is smaller than $\Delta_{\rm D}$.

One major difference between conventional 
\msr\ and \bnmr\ or \lem\  is the stopping range
of the probe. In conventional \msr\, the $\mu^+$ stopping range is
$\approx 150$ mg/cm$^2$, yielding a fraction of a mm in YBCO.
In contrast, in \bnmr\ or  \lem, 
the mean depth of the probe  can be controlled  on a  nm lengthscale from  the
surface. For implantation depths inside the superconductor, comparable or
larger than $\lambda$,  the \msr\ lineshape 
(proportional to $p(B)$) is nearly field independent
for $2B_{c1}\le B_0\le B_{c2}$ (for  $a\ll\lambda$),
and the second moment of $p(B)$ follows the formula,\cite{BrandtPRB88}
\begin{eqnarray}\label{scm-lambda}
  \sigma \approx \frac{0.00609\Phi_0}{\lambda^2(T)},
\end{eqnarray}  
neglecting the cutoff field. Using the latter makes
$\sigma$ slightly field-dependent, but the
corrections are small for fields $B_0 \ll B_{c2}$.
Outside the superconductor, 
the magnetic field inhomogeneity of the VL vanishes
over a lengthscale that depends on the spacing between vortices, $a$. In
particular,  the recovery to a uniform field
occurs on a lengthscale  of
$\frac{a}{2\pi}$.\cite{XuJMR08} The field distribution is thus strongly field
dependent when  $a(B_0)$  is of the order of
$|z|$. This is shown in Fig. \ref{simfig}(b), where $\sigma$ due to the VL
given in Eq. (\ref{scm-vl}) is plotted against the applied field at a distance
of  90 nm and 40 nm above the  surface.
In low magnetic fields,  the magnetic resonance lineshape 
outside the superconductor is sensitive to both the
intrinsic inhomogeneity of the VL as well as any 
additional broadening from disorder.  However, in high magnetic fields the
linewidth is dominated by  VL disorder. 

Taking the view that the dominant source of disorder is due to twin or grain
boundaries,\cite{NelsonPRB93} 
one can model the effect of disorder on the regular VL 
in different ways. The simplest is to assume that, in addition to 
the regular triangular lattice, a fraction of vortices is trapped
along the structural defects such as twin or 
grain boundaries as shown in the inset of Fig. \ref{simfig}(b). 
The local field in real space will be the superposition of both contributions
\begin{eqnarray}\label{bz-dis}
&&{ B}(\br,z)= { B}^{\rm vl}(\br,z) + { B}^{\rm dis}(\br,z),\nonumber\\
&&= B^{\rm vl}_0\sum_{\bk}{F}(\bk,z)e^{i\bk\cdot\br} 
+ B^{\rm dis}_0\sum_{\bg}{F}(\bg,z)e^{i\bg\cdot\br},
\end{eqnarray}
where ${ B}^{\rm vl}(\br,z)$ is the field due to the regular VL,
${ B}^{\rm dis}(\br,z)$ the field due to the vortices pinned by disorder
and $\bg$ is some generally incommensurate 
wave vector related to the pinning, which for simplicity we  take to be of the
form  $\bg= \frac{2\pi}{a'}n{\bf x}+ \frac{2\pi}{D}m{\bf y}$, where $a'$
is the spacing between vortices in a  boundary, and $D$ is the
separation between  boundaries as drawn in 
the inset of Fig. \ref{simfig}(b). The average field  is then $B_0=\sqrt{(B^{\rm
    vl}_0)^2 + (B^{\rm dis}_0)^2}$, where $B^{\rm dis}_0=fB_0$ and 
$f$ is the fraction of pinned vortices ($0\leq f<1$).
The second moment of $B(\br,z)$ from Eq. (\ref{bz-dis}) 
can  be then easily calculated:
\begin{equation}\label{scm-dvl}
\sigma^2=B^2_0\Big[ (1-f^2)\sum_{\bk\neq0} F^2(\bk,z)
+f^2\sum_{\bg\neq0} F^2(\bg,z)\Big].
\end{equation}
It is clear  from Eqs. (\ref{FourierB}) and (\ref{scm-dvl}), that the
broadening from the VL (first term) at a distance $z$ outside the
superconductor becomes  small at high magnetic fields where $|z| \gg a
$. However the  broadening outside the superconductor 
due to disorder  (second term) 
remains large provided    $|z|$ is not much larger than $\frac{D}{2\pi}$.  
Since $D$ and $f$ depend on the arrangement of twin boundaries
we expect them to be sample dependent. In addition, one may also anticipate  
that  $f$ will   decrease  at high magnetic  fields where the increased
repulsive interaction between vortices overcomes vortex-pinning.
Therefore, we assume a simplified phenomenological parameterization 
$f=\delta B_0^{-\gamma}$, where $\delta$ is temperature 
and sample dependent and $\gamma\ge 1$.

The broadening of the field distribution due to 
a regular VL can be significantly larger when introducing
the effect of disorder due to the twin and grain boundaries. 
When taking the disorder into account, $\sigma$ of Eq. (\ref{scm-dvl}) 
is no longer zero at high magnetic fields as seen in  Fig. \ref{simfig}(b).
This is because the broadening has a disorder component which decays 
on  a length scale of $D$
rather than $a$, where $D\gg a$ (we also assume $D\gg a'$ and thus ignore the
effect of the spacing within the twin boundaries).
Consequently  $\sigma$ shows a strong deviation from the ideal 
VL result as seen in Fig. \ref{simfig}-c as $D$ 
increases. In this case  the second moment from 
Eq. (\ref{scm-dvl}) no longer scales with $1/\lambda^2$ 
as predicted for an ideal VL (see Eq. (\ref{scm-lambda})). 
In particular, at low $T$, the broadening is almost $T$-independent
irrespective  of  superconducting gap structure. 
It is interesting to note that the first $\mu$SR studies on  powder samples of
cuprates showed a very flat variation in the
linewidth.\cite{HarshmanPRB87,KieflPC88}
This was taken as evidence for $s$-wave superconductivity.  Later
measurements on high quality crystals of \YBCOd\   showed a
much different low temperature behaviour,\cite{SonierPRL94} 
and, in particular, a linear variation in $1/\lambda^2(T)$ consistent with
$d$-wave pairing.\cite{HardyPRL93} 
Although the lineshapes in powders are expected to be more symmetric than in
crystals due to the additional disorder and random 
orientation,  the different temperature dependence is surprising since it was
thought that the line broadening from disorder should also scale  with
$1/\lambda^2(T)$.\cite{SonierRMP00}
 The current work provides a clear  explanation for the
discrepancy between powders and crystals. In powders, the line broadening
is  dominated by long wavelength pinning of vortices at grain
boundaries. Consequently the resulting  broadening at low temperature reflects
variations in the  vortex density and is thus only weakly dependent on
temperature. In later work on crystals, the  contribution  from such long
wavelength pinning is much less important. This is evident from bulk \msr\ in
crystals where one observes the expected characteristic lineshape associated
with a VL.\cite{SonierRMP00}

\section{Experimental details}\label{exp_section}
The measurements were carried out on three different near-optimally 
doped YBCO samples, two flux-grown single crystals and a thin film. 
I) The twinned single crystal in the 
form of a platelet $\sim0.5$ mm thick with an area
$\sim2\times3$ mm$^2$ had $T_C=92.5$ K. 
It was mechanically polished with 0.05 $\mu$m alumina, then 
chemically etched with a dilute (0.8\%) Bromine solution followed by
annealing at 200$^{\circ}$ C in dry N$_2$ to improve the surface quality.
It was then sputter coated with a 120 nm thick Ag film (99.99 \% purity) 
at room temperature in an Ar pressure of 30 mTorr. 
The deposition rate was 0.5 \AA/s, 
and to ensure Ag uniformity, the crystal was rotated.
II) The optimally doped detwinned single crystal had 
$T_C=92.5$ K, $\sim0.5$ mm thickness, and area
$\sim3\times3$ mm$^2$. The crystal  was cleaned, annealed, 
and mechanically detwinned.
A 120 nm thick Ag, from the same source as above, was sputtered 
onto the prepared surface under similar conditions.  
III) The film of $T_C=$ 87.5 K, critical current density 
$J_C=2.10^6$ A/cm$^3$ and 600 nm thickness,
supplied by THEVA (Ismaning, Germany), was grown by thermal 
co-evaporation on a LaAlO$_3$ substrate of area 9$\times$8 mm$^2$. 
The film was coated {\it in situ} with a 60 nm silver layer (99.99$\%$ purity).

The experiments were performed using the \bnmr\ spectrometer
at the ISAC facility in 
TRIUMF, Canada, where a highly nuclear-spin-polarized beam (intensity $\sim
10^6$ ions/s) of \elip\ is produced using collinear optical pumping with 
circularly polarized laser light.\cite{KieflPC03}
The beam is directed onto the sample which is mounted on the cold finger 
of a He flow cryostat and positioned in the centre of 
 a high homogeneity superconducting solenoid.  
The beamline and  entire spectrometer
are maintained in ultrahigh vacuum ($10^{-10}$ Torr).
In \bnmr\ measurements, the \Li\ nuclear spin polarization is monitored via 
its asymmetric radioactive beta decay (lifetime $\tau = 1.203$ s), where the
high energy (several MeV) beta electron is 
emitted preferentially opposite to the nuclear spin direction.
The experimental asymmetry, defined as the ratio $\frac{F-B}{F+B}$ of the
count rates in two plastic scintillation detectors 
placed in front (F) and at the back (B) of the sample,
is proportional to the probe's spin polarization.\cite{KieflPC03,MorrisPRL04}

\begin{figure}[t]
  \includegraphics[width=\columnwidth]{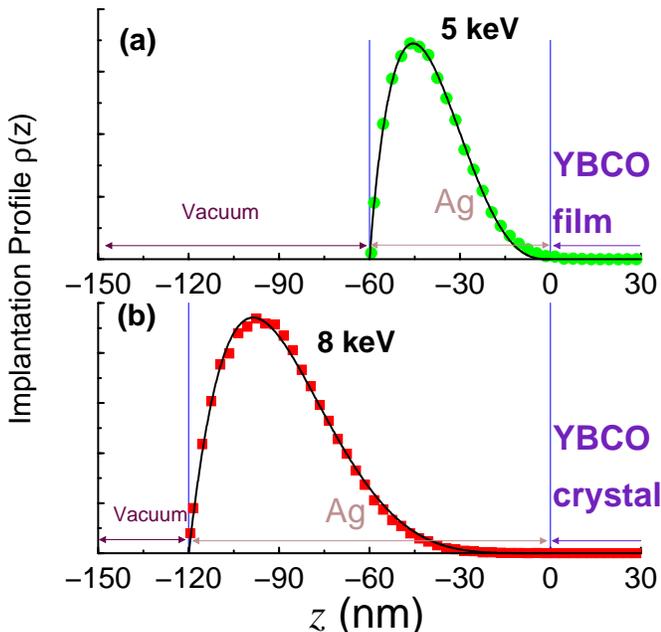}
  \caption{(Color online) Implantation profiles of \Li\ at energies of (a)
    5 keV into 60 nm of Ag with the mean at 40 nm away from YBCO film, and
    (b) 8 keV into  120 nm of Ag with 
    the mean at 90 nm away from YBCO crystals, as calculated via
    TRIM.SP.\cite{trim} Solid lines are phenomenological fits.}
  \label{trimfig}
\end{figure}

The whole spectrometer can be biased at high voltage, 
allowing one to tune the implantation energy of \Li\ ions and their 
implantation depth between 5-200 nm.
Therefore, the implanted \Li\ can monitor the depth
dependence of the local magnetic field distribution in materials at nm scale
by measuring the NMR lineshape  in a manner analogous to conventional 
NMR.\cite{SalmanPRL07,MorrisPRL04,SalmanPRB04} 
In this work, the \elip\ ions are decelerated to
stop in the Ag overlayer deposited on each of the three YBCO
samples. Implantation  profiles of \elip\ were calculated using the  TRIM.SP
code,\cite{trim} examples of which are  shown in Fig. \ref{trimfig}.  The
implantation energies used in this study (8 keV in the
crystal samples and 5 keV the film), were tuned to 
stop all the \Li\ within the Ag. 
The mean distances are 90 and 40 nm from the Ag/YBCO interface in
the crystals and film, respectively.

The \bnmr\ measurement is carried out by monitoring the time averaged nuclear 
polarization through the beta decay asymmetry, 
as a function of the radio frequency (RF) $\omega$
of a small transverse oscillating  magnetic field 
${\bf B}_1=B_1\cos(\omega  t)\hat{x}$, where $B_1\sim0.01$ mT.
When $\omega$ matches the Larmor frequency  
$\omega_{\rm Li}=\gamma_{\rm Li}B_{\rm local}$, where 
$\gamma_{\rm Li}=6.3015$ kHz/mT is the gyromagnetic ratio 
and $B_{\rm local}$ is the local field, 
the \Li\ spins precess about $B_{\rm local}$, causing a loss of polarization.
To establish the vortex state in the YBCO samples, they are cooled in a 
static magnetic fields
$B_0\ge B_{\rm c1}$ applied parallel to the $c$-axis of
YBCO (normal to the film and platelet crystals). 
$B_0$ is also parallel to both the initial 
nuclear spin polarization and the beam  direction.
The local field sensed by the \Li\ is 
determined by  the applied field and the internal
magnetic field generated by the screening currents  associated with the vortex
lattice. Thus,  $B_{\rm local}$ is distributed over a range of values, which
can be calculated using 
\equ{pB} {p(B)=\int_{-d}^{0}{\rm d}z\rho(z)\frac{1}{A}\int_{A}{\rm d}r
  \delta(B-B({\br,z})).}
where $\rho(z)$ is the implantation profile calculated using TRIM.SP
given in Fig. \ref{trimfig}.

When \elip\ is implanted in Ag (with no superconducting substrate) 
at temperatures below 100 K, it exhibits a single narrow 
resonance at the Larmor frequency.\cite{MorrisPRL04}
The resonance should yield an approximately 
Gaussian distribution caused by nuclear dipolar moments.\cite{AbragamBook}
However,  continuous wave RF leads to a
power-broadened Lorentzian lineshape, whose linewidth is small (1 kHz
$\approx 0.15$~mT) and corresponds to the dipolar broadening due to
the $^{107,109}$Ag nuclear moments and RF power
broadening.\cite{ParolinPRB08}
In the presence of any additional magnetic inhomogeneity in the Ag, 
due for example to a VL associated with a  superconducting
substrate, the observed resonance lineshape will be a convolution of the
narrow  RF power broadened Lorentzian of Ag with the (depth
dependent) field distribution due to the VL in the substrate. There are unique
aspects  of  measuring  the  field distribution in the Ag overlayer compared
to  the superconductor itself. As mentioned above, in high magnetic field, it
is possible to isolate  and  study the broadening due to VL disorder that
occurs on a long length scale.  
Also  in low magnetic fields, where the broadening is dominated by the VL,
it should be possible to  measure  $\lambda$ in magnetic
superconductors since the field distribution above the sample is free of any
internal hyperfine fields that make a bulk measurement 
impossible.\cite{SaadaouiPB09-V}

\begin{figure}[t]
  \includegraphics[width=\columnwidth]{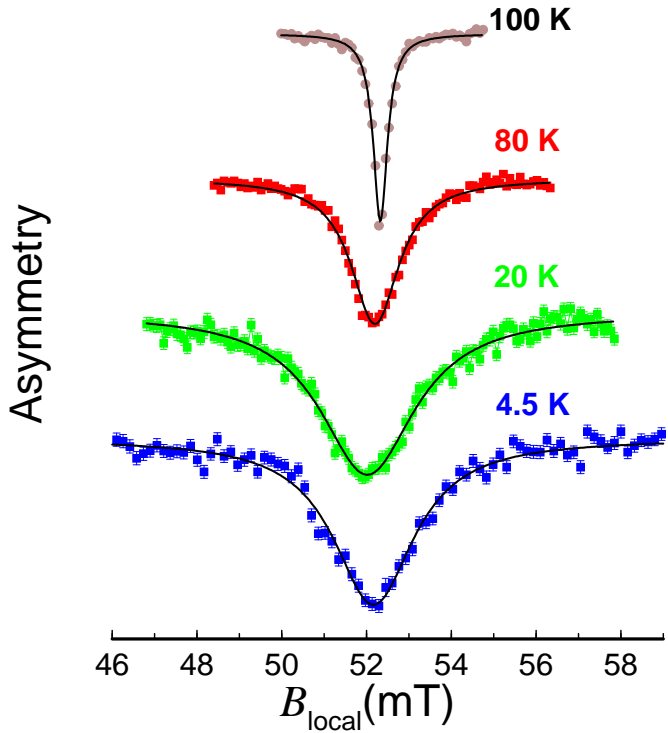}
  \caption{(Color online) 
    $\beta$NMR resonances in Ag/YBCO (crystal I) at temperatures 
    100 K, 80 K, 20 K, and 4.5 K measured  
    in a magnetic field  $B_{0}$ of $52.3$ mT
    applied along YBCO $c$-axis. The solid lines are best fits using
    a Lorentzian. }
  \label{asymfig}
\end{figure}
\begin{figure}
  \includegraphics[width=\columnwidth]{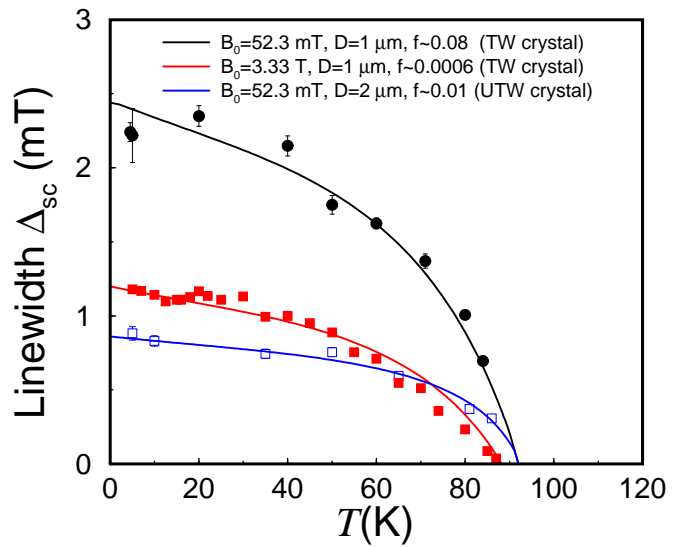}
  \caption{(Color online) The vortex-related broadening below $T_C$,
    $\Delta_{\rm sc}(T)=\Delta(T)-\Delta_{\rm ns}$ 
    of the twinned (full symbols) and detwinned 
    (opaque squares) YBCO crystals in
    an applied field $B_0$.    $\Delta(T)$ is the linewidth at
    temperature $T$ of the Lorentzian fits and $\Delta_{\rm ns}$ is the
    constant linewidth in the normal state. 
    Solid lines represent a fit using $\Delta_{\rm DVL}= 2.355\sigma$
    where $\sigma$ is given in Eq. (\ref{scm-dvl}) and
    $D$ and $f$ are varied to fit the data. A $d$-wave temperature dependence
    of     $\lambda(T)$ in YBCO is used,\cite{Bonn} 
    where $\lambda(0)=150$ nm. }
  \label{fwhmfig}
\end{figure}

\begin{figure}[t]
  \includegraphics[width=\columnwidth]{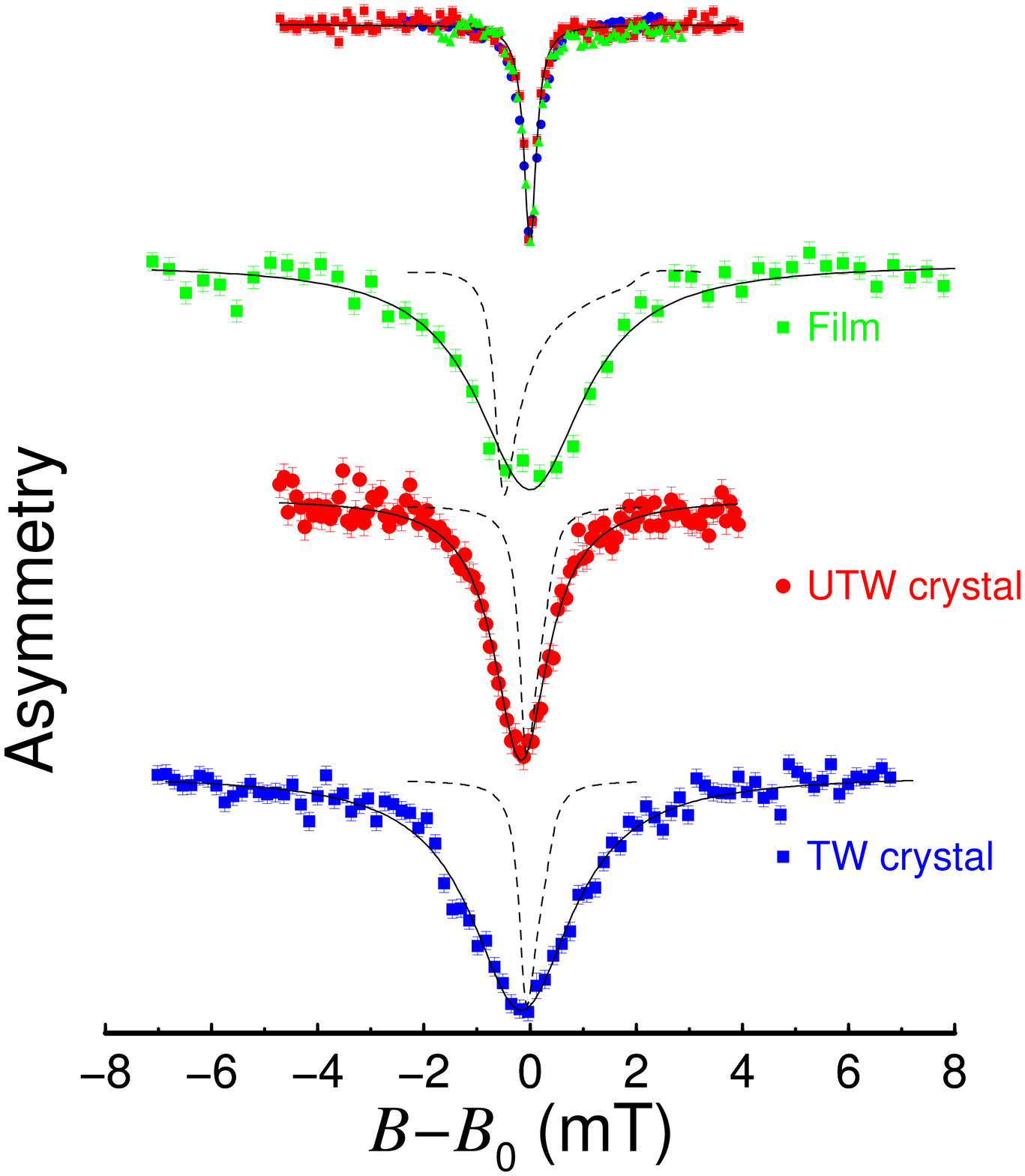}
  \caption{(Color online) Comparison of the field distributions 
    in the three samples taken at temperatures 100 K (top panel),
    5 K (crystals) and 10 K (film). The x-axis is shifted
    by ${B_{0}}$, the applied field which is  52.3 (51.7) mT for the
    crystals (film). Solid lines are Lorentzian fits and dashed lines are
    the simulation described in the text.}
  \label{ybcosfig}
\end{figure}

\section{Results}
The \bnmr\ resonances were measured as a function 
of temperature under field-cooled conditions at  fields
ranging from $B_{0}=20$ mT to 3.3 T in each one of the three samples.
Fig. \ref{asymfig}, shows typical resonance 
lineshapes at various  temperatures  in sample I
with $B_{0} = 51.7$ mT. Above  $T_C$, the line broadening is small and
temperature independent as expected from nuclear dipolar broadening.
Below $T_C$, the field distribution in the Ag overlayer broadens dramatically
from the VL in the underlying superconductor. 
Such  broadening was observed in all samples and at all magnetic fields, 
although there are significant variations  
as a function of both magnetic field and sample as discussed below.
The first thing to note is that the lineshape 
is very  symmetric  and  fits well to a simple 
Lorentzian. This is much  different from  the 
asymmetric lineshape  observed  with conventional 
\msr\ in samples similar to  I and II.\cite{SonierRMP00,SonierPRL94}
The other significant difference between the current 
results and previous bulk \msr\ 
measurements\cite{SonierPRL94,SonierPRL99,SonierRMP00,SonierPRB07}
on crystals is that  the broadening at low 
temperatures is only weakly dependent on temperature,
as may be seen by comparing the resonances at 20 K and 4.5 K. 
In contrast, the broadening from an ordered VL lattice scales with
$1/\lambda^2$  and consequently in YBCO shows 
a strong linear  $T$-dependence at low temperatures 
due to the $d$-wave superconducting order.\cite{SonierPRL94,HardyPRL93}

\begin{figure}[t]
  \includegraphics[width=\columnwidth]{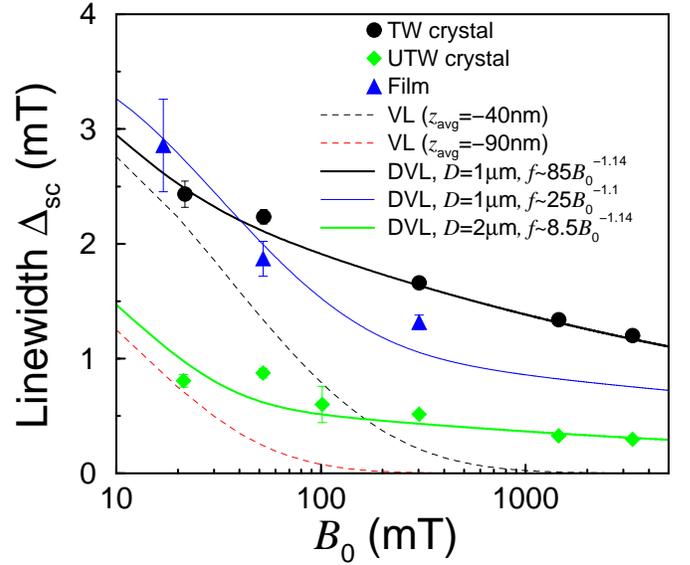}
  \caption{(Color online) Superconducting broadening $\Delta_{\rm sc}(T)$
    of the \bnmr\ resonance spectra at temperatures $\sim$ 4.5-10 K in the
    three samples. The experimental broadening is compared with the linewidth
    of an ideal VL $\Delta_{\rm VL}$ (dashed lines), and  $\Delta_{\rm DVL}$
    of a disordered VL (solid lines), which are both weighted by the \Li\
    profile given in Fig. \ref{trimfig}.}
  \label{fwhmBfig}
\end{figure}

The observed lineshape fits well to a  convolution of two
Lorentzians, one from vortices in the superconducting state 
with a full width at half maximum (FWHM)
$\Delta_{\rm sc}$, and one from  other sources determined from  
the normal state of FWHM $\Delta_{\rm ns}$. 
The width of a convolution of two Lorentzians is  
the sum of the individual widths: 
$\Delta(T) =\Delta_{\rm sc}(T) + \Delta_{\rm ns}$. 
Therefore, the contribution from vortices in the superconducting state 
can be obtained by simply subtracting  
the  temperature independent normal state width.
Fig. \ref{fwhmfig}  shows the  resulting $\Delta_{\rm sc}(T)$ as one enters 
the superconducting state in samples I and II.
At low field, the measured width ($\sim2.2$ mT) at low temperature 
is larger than expected from a regular  
VL and decreases significantly in the detwinned crystal to about $\sim0.6$ mT. 
For comparison, simulations using Eq. (\ref{pB}) and the 
\Li\ stopping profile in Fig. \ref{trimfig}(b);  
indicate that the broadening due to a regular 
VL is only  $\Delta_{\rm VL}\sim0.3$ mT.
At 3.33 T, the discrepancy between the observed width (see Fig. \ref{fwhmfig})
and that expected from a regular VL is even  more dramatic.
At this high field the vortices are spaced so closely ($a\approx 27$ nm),
that there should be no detectable broadening from a regular  VL for our
stopping depths. This can
be seen clearly from the simulation in Fig. \ref{simfig}(b), where the VL
broadening approaches  zero at high fields. In contrast, the data at 3.33 T
shows  significant broadening below $T_C$ which is therefore solely 
attributed to
vortex disorder on a long  length scale.

The temperature dependence of the broadening 
is also much weaker  than expected from a regular VL in YBCO, where 
$1/\lambda^2$ has a strong linear term due to the $d$-wave order 
parameter.\cite{SonierPRL94,HardyPRL93}
The observed temperature dependence fits well to  our model of
disorder, where $\Delta_{\rm sc}(T)$ is compared to  
an estimate of the FWHM given by 
$\Delta_{\rm DVL}\approx 2.355\sigma$, 
where $\sigma$ is given in Eq. (\ref{scm-dvl}).
This leads to an estimate of $D$ of the order of a micron,
consistent with the the separation between 
twin boundaries or grain boundaries.\cite{DolanPRL89}
In the detwinned crystal, $D$ is found to be 
larger but not infinite since the detwinning is not complete.
The fraction of vortices $f$ pinned by structural 
defects in the twinned crystal is about $\sim 0.1$ at low field (52 mT) 
and decreases considerably at high field (3.33 T). 
Thus, the amplitude of the enhanced vortex density at the defects, $fB_0$, 
varies between 2-4 mT at all fields. In the detwinned crystal, $f\sim 0.01$, 
is an order of magnitude smaller than in the twinned crystal at the same
field,  with small variation in the vortex density (0.5 mT) compared
to the twinned crystal.  These results are consistent with expectations from
pinning at  twin boundaries.  In particular, one  expects the fraction of
vortices pinned  will  decrease  in the partially detwinned crystals.
Also, it is reasonable to expect that  in a  high magnetic fields the fraction
of vortices pinned will decrease due to  the smaller separation between
vortices  and resulting increase in the repulsive interaction.

In Fig. \ref{ybcosfig}, the spectra in all three samples 
above and below  $T_C$ are compared with the corresponding simulated field
distributions. The observed lineshapes  
are all symmetric, and significantly broader
than expected, showing little or no sign 
of the characteristic VL field distribution.
Simulation of the VL lineshapes (dashed lines) was done 
using Eqs. (\ref{FourierB}), (\ref{Bzr_eq}), and
(\ref{pB}), for $\lambda=150$ nm, and was convoluted with a Lorentzian
representing the normal state spectra with ${\Delta_{\rm ns}}=0.3$ mT. 
The theoretical lineshape for the film is
broader and asymmetric because it is weighted by the 
\Li\ stopping distribution  which was on average closer to the
superconductor. The lineshapes for the crystals
are almost symmetric as the ideal VL lineshape at the depths of 
an average 90 nm away from the superconductor are narrower than  
the Lorentzian they are convoluted with.
The magnetic field dependence of the superconducting linewidth 
$\Delta_{\rm sc}(T)$ at low temperatures $\sim$ 4.5-10 K 
is plotted in Fig. \ref{fwhmBfig}. 
In all samples, we find that the  broadening  is largest at low field and 
decreases gradually with increasing field.  
Also, in all cases the broadening remains large 
and well above the prediction from
a regular VL, approximated by $\Delta_{\rm VL}\approx2.355\sigma$
where $\sigma$ of an ideal VL is given in Eq. (\ref{scm-vl}) and 
weighted by the \Li\ profile given in Fig. \ref{trimfig}. 
The broadening is  substantially
smaller in the detwinned crystal compared to the other samples.
One can account for all of the data using a linewidth due to 
a disordered VL, $\Delta_{\rm DVL}\approx2.355\sigma$, where 
$\sigma$ is now given in Eq. (\ref{scm-dvl})) and 
weighted by the \Li\ profile plotted in Fig. \ref{trimfig}. 
The data is well fitted (see Fig. \ref{fwhmBfig}) 
by assuming that the twin/grain boundaries
spacing $D$ is sample dependent of the order of a few microns, and by
assuming  a phenomenological form for the fraction of pinned vortices
$f=\delta B_0^{-\gamma}$ with  $\gamma\sim 1.1$ and $\delta$  sample dependent.

\section{Discussion and Conclusions}
It is clear that the \bnmr\ lineshapes in the
Ag overlayer differ substantially from that expected from a well-ordered 
VL field distribution.  This has little  to do with the method of
observation. For example, in  the conventional superconductor NbSe$_2$,
\bnmr\ shows the expected VL lineshape.\cite{SalmanPRL07}
The lineshapes reported here in the YBCO 
film are also qualitatively  different  than that seen  with LE-$\mu$SR
in a YBCO film  coated with a 60 nm thick Ag layer.\cite{NiedermayerPRL99} 
In that experiment  the authors  found a more asymmetric lineshape in the Ag
overlayer   which was closer to that of a regular VL.
Some of this difference may be due to  the  different pinning characteristics
of the samples,  although   the YBCO film used by Niedermayer 
\etal\ was from the same source as sample III. 
Also, the LE-\msr\  experiment was probing the 
VL closer to the interface and in a lower applied field where the disorder is
less important compared to the contribution from the ordered VL.  
The symmetry and large broadening of the lineshape 
at low fields cannot be 
accounted for by VL melting (at a reentrant
vortex liquid state near $B_{c1}$) which would instead yield a motional 
narrowing of the field distribution.\cite{LeePRL93}

The observed resonances in the current experiment are dominated by  long range
variations of the vortex  density across the face of the
sample.\cite{Maggio-Aprile97} Such disorder in the VL can produce a symmetric 
lineshape,\cite{HarshmanPRB93,DivakarPRL04} and can broaden the
field distribution significantly compared to that of the corresponding ordered
state.\cite{MinkinPSS04} Weak random pinning or 
point-like disorder due to oxygen deficiency may slightly distort the VL, and 
may also broaden the lineshape.\cite{BrandtPRB88} 
However, the correlated disorder due to the
twin and grain boundaries is dominant at long wavelengths,
\cite{NelsonPRB93,OlivePRB98} 
and therefore we are mostly sensitive to the twin/grain boundaries.
 Indeed, the position of the probe outside the
superconductor  enhances its  sensitivity to long wavelength disorder,
as the proximal fields fall off with distance as $\exp(- 2\pi |z|/D)$
where $D$ is the wavelength of the inhomogeneity  of the
field.\cite{BontempsPRB91} The broadening is reduced in a 
detwinned crystal where the twin boundaries are more sparse as shown in
Fig. \ref{ybcosfig},  thus the vortex density variation across the face of the
sample is smaller than in the twinned crystals as $f$ is largely reduced.

The extrinsic broadening due to disorder at low temperature, 
$\Delta_{\rm D}=\Delta_{\rm  sc}-\Delta_{\rm VL}$,
reported  here is between 0.5 mT and 2.5 mT. This is
remarkably close to the additional Gaussian broadening required to explain
lineshapes in bulk \msr\ measurements  
on crystals.\cite{SonierRMP00,HarshmanPRB93,RisemanPRB95}
In the bulk, this extrinsic broadening is small compared to 
the intrinsic VL broadening, whereas  outside 
the sample the reverse is true. It is important to note  
the $T$-dependence of the extrinsic  broadening in Fig. \ref{fwhmfig}
{\it does not follow the superfluid density} ($\propto 1/\lambda^2$) 
which varies linearly  at low-$T$ because of the $d$-wave order
parameter.\cite{HardyPRL93,UemuraPRL89} Instead, we observe a much weaker
$T$-dependence. This is expected from our model of disorder  which occurs  on
a long length scale $D$. For example, at low-$T$ where $\lambda$ is short 
compared to $D$,  the flux density outside the sample is determined solely by 
inhomogeneities in the  vortex density, and is 
independent of the superfluid density
as the vortices are static and well-pinned in the twin boundaries. 
The current  results may
also explain early \msr\  work on HTSC powders and sintered samples which
mistakenly indicated  an $s$-wave $T$-dependence of 
$1/\lambda^2$.\cite{HarshmanPRB87,KieflPC88}  It is likely  in
these cases   the linewidth was dominated by extrinsic  VL disorder on a long
length scale. This tends to  flatten the $T$-dependence of the
linewidth and the effective $\lambda$ obtained from the
analysis.\cite{PumpinPRB90,HarshmanPRB87}  Therefore we conclude that 
although the linewidth obtained from powders can be  useful 
in making rough estimates of 
$\lambda$,  one cannot extract accurate measurements of  
$\lambda$ or its $T$-dependence without additional information 
about the source of broadening  and in particular VL disorder. 

In conclusion, we have measured the magnetic field distributions
due to the vortex state of YBCO    using \bnmr. 
We find a significant inhomogeneous broadening of the NMR attributed to the 
underlying VL in YBCO. However, the observed resonances have
several unexpected properties. In particular, they are broader and  more
symmetric than for an ideal VL.  The anomalous broadening is most evident in
high field where there is no significant contribution from the regular
VL. These effects are attributed to long wavelength disorder from  pinning at
twin or grain boundaries. The temperature dependence of the 
disorder-related broadening does not scale with  $1/\lambda^2$, suggesting
there is a contribution to the linewidth in the bulk of the vortex state
that does not track the superfluid density. This is likely to have only  a
minor effect on the interpretation of data on crystals where the observed
lineshape  is close to that expected from a well  ordered VL. However,
it can be significant in powders or crystals where there is substantial
disorder in the VL. In particular,  when the broadening is dominated by VL
disorder on a long length scale  (i.e. much bigger than $\lambda$)   the
temperature dependence of the linewidth does not scale with $1/\lambda^2$ and
therefore cannot be used to determine the symmetry of the superconducting gap. 

\begin{acknowledgments}
  We would like to thank D. A. Bonn, W. N. Hardy, and R. Liang for providing 
  the YBCO crystals. We would also like to acknowledge R. Abasalti, 
  D. Arseneau, and S. Daviel for expert technical support, 
  and NSERC, CIFAR for financial support.
\end{acknowledgments}


\end{document}